\documentstyle[12pt]{article}

\newcommand \beq{\begin{eqnarray}}
\newcommand \eeq{\end{eqnarray}}

\catcode`\@=11

\def\ps@myheadings{\let\@mkboth\@gobbletwo
\def\@oddhead{\hbox{} 
\rightmark\hfil\ninerm\thepage}
\def\@oddfoot{}\def\@evenhead{\ninerm\thepage\hfil 
\leftmark\hbox{}}\def\@evenfoot{}
\def\sectionmark##1{}\def\subsectionmark##1{}}

\evensidemargin
0.30truein
\begin{document}

\begin{titlepage}
\begin{flushright}

{Saclay-T93/122}
\end{flushright}
\vspace*{3cm}
\begin{center}
\baselineskip=13pt
{\Large SOFT FIELDS AND HARD PARTICLES IN\\ HOT GAUGE PLASMAS\\}
\vspace{1cm}
 Edmond IANCU\\

{\it Service de Physique Th\'eorique
\footnote{Laboratoire de la Direction des
Sciences de la Mati\`ere du Commissariat \`a l'Energie
Atomique}, CE-Saclay}\\
\baselineskip=12pt
{\it  91191 Gif-sur-Yvette, France}\\
\vskip0.5cm
October 1993
\end{center}

\vskip 2cm \begin{abstract}
It is shown that the long wavelength excitations of a quark-gluon plasma  may
 be described as collective oscillations of  self-consistent average
 fields to which the plasma particles couple. Their properties are
obtained from  a  set of coupled mean field and kinetic equations,
derived from  the general Dyson-Schwinger equations, as
the leading order in a systematic expansion in powers of the  coupling.
 By solving the kinetic equations,
 one obtains in closed form the generating functionals
 for all the leading order amplitudes between soft quasiparticles,
the so-called  ``hard thermal loops''.
\end{abstract}
\vskip 1cm

\begin{flushleft}
Talk given at the 3rd Workshop on Thermal Field Theories and
their Applications, \\
August 15--27, 1993, Banff, Alberta, Canada

\end{flushleft}
\end{titlepage}

\newcommand{\symbolfootnote}{\renewcommand{\thefootnote}
        {\fnsymbol{footnote}}}
\renewcommand{\thefootnote}{\fnsymbol{footnote}}
\newcommand{\alphfootnote}
        {\setcounter{footnote}{0}
         \renewcommand{\thefootnote}{\sevenrm\alph{footnote}}}

\newcounter{sectionc}\newcounter{subsectionc}\newcounter{subsubsectionc}
\renewcommand{\section}[1] {\vspace{0.6cm}\addtocounter{sectionc}{1}
\setcounter{subsectionc}{0}\setcounter{subsubsectionc}{0}\noindent
        {\bf\thesectionc. #1}\par\vspace{0.4cm}}
\renewcommand{\subsection}[1] {\vspace{0.6cm}\addtocounter{subsectionc}{1}
        \setcounter{subsubsectionc}{0}\noindent
        {\it\thesectionc.\thesubsectionc. #1}\par\vspace{0.4cm}}
\renewcommand{\subsubsection}[1]
{\vspace{0.6cm}\addtocounter{subsubsectionc}{1}
        \noindent {\rm\thesectionc.\thesubsectionc.\thesubsubsectionc.
        #1}\par\vspace{0.4cm}}
\newcommand{\nonumsection}[1] {\vspace{0.6cm}\noindent{\bf #1}
        \par\vspace{0.4cm}}

\newcounter{appendixc}
\newcounter{subappendixc}[appendixc]
\newcounter{subsubappendixc}[subappendixc]
\renewcommand{\thesubappendixc}{\Alph{appendixc}.\arabic{subappendixc}}
\renewcommand{\thesubsubappendixc}
        {\Alph{appendixc}.\arabic{subappendixc}.\arabic{subsubappendixc}}

\renewcommand{\appendix}[1] {\vspace{0.6cm}
        \refstepcounter{appendixc}
        \setcounter{figure}{0}
        \setcounter{table}{0}
        \setcounter{equation}{0}
        \renewcommand{\thefigure}{\Alph{appendixc}.\arabic{figure}}
        \renewcommand{\thetable}{\Alph{appendixc}.\arabic{table}}
        \renewcommand{\theappendixc}{\Alph{appendixc}}
        \renewcommand{\theequation}{\Alph{appendixc}.\arabic{equation}}
        \noindent{\bf Appendix \theappendixc #1}\par\vspace{0.4cm}}
\newcommand{\subappendix}[1] {\vspace{0.6cm}
        \refstepcounter{subappendixc}
        \noindent{\bf Appendix \thesubappendixc. #1}\par\vspace{0.4cm}}
\newcommand{\subsubappendix}[1] {\vspace{0.6cm}
        \refstepcounter{subsubappendixc}
        \noindent{\it Appendix \thesubsubappendixc. #1}
        \par\vspace{0.4cm}}

\def\abstracts#1{{
        \centering{\begin{minipage}{30pc}\tenrm\baselineskip=12pt\noindent
        \centerline{\tenrm ABSTRACT}\vspace{0.3cm}
        \parindent=0pt #1
        \end{minipage} }\par}}

\newcommand{\bibit}{\it}
\newcommand{\bibbf}{\bf}
\renewenvironment{thebibliography}[1]
        {\begin{list}{\arabic{enumi}.}
        {\usecounter{enumi}\setlength{\parsep}{0pt}
\setlength{\leftmargin 1.25cm}{\rightmargin 0pt}
         \setlength{\itemsep}{0pt} \settowidth
        {\labelwidth}{#1.}\sloppy}}{\end{list}}

\topsep=0in\parsep=0in\itemsep=0in
\parindent=1.5pc

\newcounter{itemlistc}
\newcounter{romanlistc}
\newcounter{alphlistc}
\newcounter{arabiclistc}
\newenvironment{itemlist}
        {\setcounter{itemlistc}{0}
         \begin{list}{$\bullet$}
        {\usecounter{itemlistc}
         \setlength{\parsep}{0pt}
         \setlength{\itemsep}{0pt}}}{\end{list}}

\newenvironment{romanlist}
        {\setcounter{romanlistc}{0}
         \begin{list}{$($\roman{romanlistc}$)$}
        {\usecounter{romanlistc}
         \setlength{\parsep}{0pt}
         \setlength{\itemsep}{0pt}}}{\end{list}}

\newenvironment{alphlist}
        {\setcounter{alphlistc}{0}
         \begin{list}{$($\alph{alphlistc}$)$}
        {\usecounter{alphlistc}
         \setlength{\parsep}{0pt}
         \setlength{\itemsep}{0pt}}}{\end{list}}

\newenvironment{arabiclist}
        {\setcounter{arabiclistc}{0}
         \begin{list}{\arabic{arabiclistc}}
        {\usecounter{arabiclistc}
         \setlength{\parsep}{0pt}
         \setlength{\itemsep}{0pt}}}{\end{list}}

\newcommand{\fcaption}[1]{
        \refstepcounter{figure}
        \setbox\@tempboxa = \hbox{\tenrm Fig.~\thefigure. #1}
        \ifdim \wd\@tempboxa > 6in
           {\begin{center}
        \parbox{6in}{\tenrm\baselineskip=12pt Fig.~\thefigure. #1 }
            \end{center}}
        \else
             {\begin{center}
             {\tenrm Fig.~\thefigure. #1}
              \end{center}}
        \fi}

\newcommand{\tcaption}[1]{
        \refstepcounter{table}
        \setbox\@tempboxa = \hbox{\tenrm Table~\thetable. #1}
        \ifdim \wd\@tempboxa > 6in
           {\begin{center}
        \parbox{6in}{\tenrm\baselineskip=12pt Table~\thetable. #1 }
            \end{center}}
        \else
             {\begin{center}
             {\tenrm Table~\thetable. #1}
              \end{center}}
        \fi}

\def\@citex[#1]#2{\if@filesw\immediate\write\@auxout
        {\string\citation{#2}}\fi
\def\@citea{}\@cite{\@for\@citeb:=#2\do
        {\@citea\def\@citea{,}\@ifundefined
        {b@\@citeb}{{\bf ?}\@warning
        {Citation `\@citeb' on page \thepage \space undefined}}
        {\csname b@\@citeb\endcsname}}}{#1}}

\newif\if@cghi
\def\cite{\@cghitrue\@ifnextchar [{\@tempswatrue
        \@citex}{\@tempswafalse\@citex[]}}
\def\citelow{\@cghifalse\@ifnextchar [{\@tempswatrue
        \@citex}{\@tempswafalse\@citex[]}}
\def\@cite#1#2{{$\null^{#1}$\if@tempswa\typeout
        {IJCGA warning: optional citation argument
        ignored: `#2'} \fi}}
\newcommand{\citeup}{\cite}

\def\fnm#1{$^{\mbox{\scriptsize #1}}$}
\def\fnt#1#2{\footnotetext{\kern-.3em
        {$^{\mbox{\sevenrm #1}}$}{#2}}}

\font\twelvebf=cmbx10 scaled\magstep 1
\font\twelverm=cmr10 scaled\magstep 1
\font\twelveit=cmti10 scaled\magstep 1
\font\elevenbfit=cmbxti10 scaled\magstephalf
\font\elevenbf=cmbx10 scaled\magstephalf
\font\elevenrm=cmr10 scaled\magstephalf
\font\elevenit=cmti10 scaled\magstephalf
\font\bfit=cmbxti10
\font\tenbf=cmbx10
\font\tenrm=cmr10
\font\tenit=cmti10
\font\ninebf=cmbx9
\font\ninerm=cmr9
\font\nineit=cmti9
\font\eightbf=cmbx8
\font\eightrm=cmr8
\font\eightit=cmti8

\centerline{\tenbf SOFT FIELDS AND HARD PARTICLES IN HOT GAUGE PLASMAS}
\baselineskip=16pt
\vspace{0.8cm}
\centerline{\tenrm Edmond IANCU}
\baselineskip=13pt
\centerline{\tenit Service de Physique Th\'eorique
\footnote{Laboratoire de la Direction des
Sciences de la Mati\`ere du Commissariat \`a l'Energie
Atomique}, CE-Saclay }
\baselineskip=12pt
\centerline{\tenit  91191 Gif-sur-Yvette, France}
\vspace{0.9cm}
\abstracts{
It is shown that the long wavelength excitations of a quark-gluon plasma  may
 be described as collective oscillations of  self-consistent average
 fields to which the plasma particles couple. Their properties are
obtained from  a  set of coupled mean field and kinetic equations,
derived from  the general Dyson-Schwinger equations, as
the leading order in a systematic expansion in powers of the  coupling.
 By solving the kinetic equations,
 one obtains in closed form the generating functionals
 for all the leading order amplitudes between soft quasiparticles,
the so-called  ``hard thermal loops''.
}
\vfill
\twelverm   
\baselineskip=14pt


\def\square{\hbox{{$\sqcup$}\llap{$\sqcap$}}}   
\def\grad{\nabla}                               
\def\del{\partial}                              

\def\frac#1#2{{#1 \over #2}}
\def\smallfrac#1#2{{\scriptstyle {#1 \over #2}}}
\def\half{\ifinner {\scriptstyle {1 \over 2}}
   \else {1 \over 2} \fi}

\def\bra#1{\langle#1\vert}              
\def\ket#1{\vert#1\rangle}              

\def\simge{\mathrel{%
   \rlap{\raise 0.511ex \hbox{$>$}}{\lower 0.511ex \hbox{$\sim$}}}}
\def\simle{\mathrel{
   \rlap{\raise 0.511ex \hbox{$<$}}{\lower 0.511ex \hbox{$\sim$}}}}


\def\parenbar#1{{\null\!                        
   \mathop#1\limits^{\hbox{\fiverm (--)}}       
   \!\null}}                                    
\def\nunubar{\parenbar{\nu}}
\def\ppbar{\parenbar{p}}


\def\buildchar#1#2#3{{\null\!                   
   \mathop#1\limits^{#2}_{#3}                   
   \!\null}}                                    
\def\overcirc#1{\buildchar{#1}{\circ}{}}


\def\slashchar#1{\setbox0=\hbox{$#1$}           
   \dimen0=\wd0                                 
   \setbox1=\hbox{/} \dimen1=\wd1               
   \ifdim\dimen0>\dimen1                        
      \rlap{\hbox to \dimen0{\hfil/\hfil}}      
      #1                                        
   \else                                        
      \rlap{\hbox to \dimen1{\hfil$#1$\hfil}}   
      /                                         
   \fi}                                         %


\def\subrightarrow#1{
  \setbox0=\hbox{
    $\displaystyle\mathop{}
    \limits_{#1}$}
  \dimen0=\wd0
  \advance \dimen0 by .5em
  \mathrel{
    \mathop{\hbox to \dimen0{\rightarrowfill}}
       \limits_{#1}}}                           

\def\real{\mathop{\rm Re}\nolimits}     
\def\imag{\mathop{\rm Im}\nolimits}     

\def\tr{\mathop{\rm tr}\nolimits}       
\def\Tr{\mathop{\rm Tr}\nolimits}       
\def\Det{\mathop{\rm Det}\nolimits}     

\def\mod{\mathop{\rm mod}\nolimits}     
\def\wrt{\mathop{\rm wrt}\nolimits}     


\def\TeV{{\rm TeV}}                     
\def\GeV{{\rm GeV}}                     
\def\MeV{{\rm MeV}}                     
\def\KeV{{\rm KeV}}                     
\def\eV{{\rm eV}}                       

\def\mb{{\rm mb}}                       
\def\mub{\hbox{$\mu$b}}                 
\def\nb{{\rm nb}}                       
\def\pb{{\rm pb}}                       

%
\def\journal#1#2#3#4{\ {#1}{\bf #2} ({#3})\  {#4}}

\def\AdvPhys{\journal{Adv.\ Phys.}}
\def\AnnPhys{\journal{Ann.\ Phys.}}
\def\EurophysLett{\journal{Europhys.\ Lett.}}
\def\JApplPhys{\journal{J.\ Appl.\ Phys.}}
\def\JMathPhys{\journal{J.\ Math.\ Phys.}}
\def\LettNuovoCimento{\journal{Lett.\ Nuovo Cimento}}
\def\Nature{\journal{Nature}}
\def\NPA{\journal{Nucl.\ Phys.\ {\bf A}}}
\def\NPB{\journal{\it {Nucl.\ Phys.\ {\bf B}}}}
\def\NuovoCimento{\journal{Nuovo Cimento}}
\def\Physica{\journal{Physica}}
\def\PLA{\journal{Phys.\ Lett.\ {\bf A}}}
\def\PLB{\journal{Phys.\ Lett.\ {\bf B}}}
\def\PhysRev{\journal{Phys.\ Rev.}}
\def\PRC{\journal{Phys.\ Rev.\ {\bf C}}}
\def\PRD{\journal{\it {Phys.\ Rev.\ {\bf D}}}}
\def\PRL{\journal{\it {Phys.\ Rev.\ Lett.}}}
\def\PhysRept{\journal{Phys.\ Repts.}}
\def\ProcNatlAcadSci{\journal{Proc.\ Natl.\ Acad.\ Sci.}}
\def\ProcRoySoc{\journal{Proc.\ Roy.\ Soc.\ London Ser.\ A}}
\def\RevModPhys{\journal{Rev.\ Mod.\ Phys.}}
\def\Science{\journal{Science}}
\def\SovPhysJETP{\journal{\it {Sov.\ Phys.\ JETP}}}
\def\SovPhysJETPLett{\journal{Sov.\ Phys.\ JETP Lett.}}
\def\SovJNuclPhys{\journal{\it {Sov.\ J.\ Nucl.\ Phys.}}}
\def\SovPhysDoklady{\journal{Sov.\ Phys.\ Doklady}}
\def\ZPhys{\journal{Z.\ Phys.}}
\def\ZPhysA{\journal{Z.\ Phys.\ A}}
\def\ZPhysB{\journal{Z.\ Phys.\ B}}
\def\ZPhysC{\journal{Z.\ Phys.\ C}}


\setcounter{equation}{0}
\parindent=20pt

\section{Introduction}
The present talk is concerned with some recent progress  in the
understanding of the collective modes propagating in a gauge plasma at high
temperature. The results to be reported here offer a consistent
and gauge covariant description for the off-equilibrium dynamics of
a quark-gluon (or electron-photon) plasma, valid at leading order in the
gauge coupling constant $g$. (I consider QCD at a temperature $T$ well into
its deconfined phase and assume a perturbative regime, $g\ll 1$.)
Such a description may be viewed as a first step towards a more complete
treatement of the plasma transport and relaxation properties. Besides,
as we shall see, it is also relevant for a physical understanding
 of the newly developed  resummation schemes,
allowing for consistent  computations in hot gauge theories\cite{Pisarski}.

As it is clear by now,  in high-$T$ field theory
 the naive loop expansion fails. As shown by
Braaten and Pisarski\cite{Pisarski}, one-loop corrections
to amplitudes with {\it soft} external momenta $P_1,...,P_n$, (i.e., such
that $P_j\simle gT$ for any $j$) are as important as the corresponding tree
level amplitudes and, correspondingly, have to be resummed in consistent
 higher order computations\cite{Pisarski}. The
 dominant one-loop contributions are
called ``hard thermal loops''(HTL), as they arise from integrating over
{\it hard} internal momenta $k$, (i.e., over $k$ of order $T$).
Isolated first  in the 2-point amplitudes\cite{Klimov},
 and, later on, in general $n$-point amplitudes\cite{Frenkel,Pisarski}
($n\ge 2$), the HTL's show some outstanding properties,
which are left largely unexplained by their original derivation
in terms of Feynman diagrams. I summarize here some of them:

(a) There are some systematic ``selection''  rules in the appearance
of the HTL's:
\hspace*{1.4cm}-- no HTL with more then one pair of fermion lines;\\
\hspace*{1.4cm}-- in QED: no HTL with more then 2 external photons;\\
\hspace*{1.4cm}-- in QCD: HTL's in all multi-gluon amplitudes;\\
\hspace*{3.5cm}no HTL with ghost external lines.

(b) The HTL's present a simple structure. E.g.,  the HTL for the
 gluon polarization tensor is identical to the dielectric
tensor  for a {\it classical} ultrarelativistic plasma, as already computed
by Silin in 1960 from kinetic theory\cite{Silin}.

(c) HTL's are generally {\it momentum dependent} through ``Vlasov-like'' (or
 ``eikonal-like'') denominators, of the  type  $1\,/(v\cdot P)$, where
 $v^\mu\,=\,(1,\,{\vec v})$ is the loop velocity, while $P$ is a linear
combination of the external momenta.

(d) HTL's are independent of the gauge fixing condition used in their
evaluation.

(e) HTL's obey {\it ``abelian-like'' Ward identities} even in the case of
QCD. This is equivalent to a statement of gauge
invariance and suggested the construction of
 an effective action generating all the HTL's\cite{Wong,Braaten92}.
 This is a nonlocal functional of the soft fields,
whose properties are further investigated in various recent works\cite{Weldon}.

(f) As already notified, the HTL's
have to be resummed, for consistency,
 into (non-local) {\it effective propagators and vertices}.
By using them in perturbative computations, one has obtained
gauge independent results for the damping of the collective
modes\cite{Pisarski},
as well as sensible values for some transport coefficients\cite{Baym}.

The need for resummation in the perturbative expansion,
as well as the simple properties above, suggest some {\it simple
underlying physics}, whose understanding was the main motivation
of the present work. To this aim, a new approach to the problem
 has been initiated by  Jean-Paul Blaizot and myself,
 by working directly at the level of the
equations of motion rather than on the Feynman diagrams\cite{us}.
Our results, to be discussed below, may be
summarized in the following {\it physical picture}:

The HTL's describe  polarization properties  related to
     {\it the ``semiclassical'' collective dynamics of the hard particles} in
 {\it selfconsistent mean fields.} Resummation is
 necessary to properly account for {\it the collective character of
 the soft excitations} in higher orders of perturbation theory.

 In order to see how this picture emerges, let me start by recalling
 some typical energy and length scales in the ultrarelativistic plasma.
In thermal equilibrium, at high $T$,
quarks and gluons have typical energies and momenta of
order $T$, and will be referred to as ``hard'' particles.
Their number density is proportional to $T^3$, and therefore
the typical interparticle distance $r_0$ is $\sim 1/T$.
When coupled to  slowly varying disturbances, the plasma may develop
a {\it collective behaviour} on a typical length/time scale of order $1/gT$. A
familiar example is Debye's screening.
Collectivity arises because any motion taking place over a distance scale
$\lambda\sim 1/gT$ may involve coherently a
 large number of hard particles. This gives rise to long
 range fluctuations in the densities of the associated
quantum numbers (like charge, color, spin, baryonic nb.),
which are conveniently described in terms of long wavelength oscillations
in the {\it selfconsistent mean fields} carrying the appropriate quantum
 nbs.: $\langle A_\mu(X)\rangle$,
$\langle \psi(X)\rangle$,  $\langle \bar\psi(X)\rangle$, ...
Such fields, to be referred as ``soft'' in what follows ($\lambda\sim 1/gT$),
 represent  the average motion of the hard particles over distances
 large compared to their mean separation ($\lambda \gg r_0$).

Our main objective is to  characterize the properties
of the collective modes (i.e., their propagation and mutual
interactions) {\it at leading order in $g$}.
As we shall see shortly, these properties are essentially
 determined by {\it the
interactions  between the soft fields and the hard thermal particles},
which explains the title of this talk. In particular, the HTL's  play a
decisive role in this description, and all their remarkable
 properties alluded to before  find in this context a natural explanation.

\section{An approximation scheme with three facets}


I shall consider the response of the plasma to weak and
slowly varying external disturbances.
When studying off-equilibrium perturbations, one is usually facing three kinds
of approximations, involving, respectively, the strength of the coupling, the
amplitude of the field oscillations, and the wavelength or period of the
 modes that one is studying.
However, for the ultrarelativistic plasma, it turns out that {\it all three
approximations are controlled by the single small parameter, $g$}.
They should be viewed, then, as three facets of the same approximation, namely
 an expansion in powers of $g$.
The naive loop expansion fails because it takes
care only of the first facet,
 namely the expansion in powers of the interaction vertices.
To understand how the coupling constant $g$ controls the mean fields aspects
of this problem, recall that

(a) the mean fields are {\it slowly varying}, $\lambda\sim 1/gT$,
i.e. they carry a typical momentum $P\sim gT$. Then {\it
 any soft derivative carries a power of $g$}:
$i\del\langle\phi(x)\rangle\sim gT\langle\phi(x)\rangle$, where
$\langle\phi(x)\rangle$ denotes any of the average fields
$\langle A_\mu(x)\rangle,\,\langle\psi(x)\rangle, ...$

(b) the mean fields are {\it weak (plasma close to equilibrium)}. This
 puts  constraints on the strength of the mean fields which also depend on
 $g$. Indeed, the plasma will be weakly perturbed
if, for instance, the color force $\sim g\langle F\rangle$ acting on a
typical particle with momentum $k\sim T$, produces a small change in its
 momentum, i.e. $\Delta k=g\langle F\rangle\Delta t\ll k$;
 (here, $\langle F_{\mu\nu}\rangle$ is the gauge field strength
tensor). One sees that, for $\Delta t\sim \lambda\sim 1/gT$ and
 $\Delta k\simle gT$, we have $\langle F\rangle\simle gT^2$. More
generally,  there exists a limit in which all terms in our final
equations of motion (\ref{avpsi})--(\ref{N}) are of the same order of
 magnitude.  This is achieved when $\langle F\rangle\sim gT^2$
and $\langle\bar\psi\rangle
\langle\psi\rangle\sim gT^3$, i.e.,  {\it $g$ times a
power of $T$  equal to their canonical dimension}.
In this limit, $g\langle A_\mu\rangle\sim gT$
is of the same order as the derivative of a ``slowly varying''
quantity,
$i\del_\mu\langle \phi\rangle\sim g\langle A_\mu\rangle\,\langle\phi\rangle$,
which insures the consistency of the soft
 covariant derivatives
$D_\mu=\del_\mu+ig\langle A_\mu\rangle$.
In what follows,  I shall consider
this limit, where the equations keep their full non-abelian structure.

Our strategy is by now clear: we have to expand
 the exact (real-time) Dyson-Schwinger
 equations of motion in powers of $g$ and to
 preserve consistently the leading, nontrivial, terms.
In applying this procedure, we find that the  dominant
interactions which determine  the response of the plasma  are those which take
place between the hard particles and  the soft mean fields.
Collisions between hard particles may be ignored in this order,
which allows us to truncate  the hierarchy at the level
of the 2-point functions. That is, we end up with
a {\it coupled system of equations for the soft mean fields and for
the hard particles 2-point functions}, to be presented  now.

\setcounter{equation}{0}
\section{Mean field and kinetic equations}

The equations for the soft mean fields  are
\beq
\label{avpsi}
i\slashchar{D}\, \psi(X)=\eta(X)+\eta^{ind}(X),
\eeq
\beq
\label{avA}
\left [\, D^\nu,\, F_{\nu\mu}(X)\,\right ]^a
-g  \bar\psi (X)\gamma_\mu t^a \psi(X)
=\,j_\mu^a(X)+j_\mu^{ind\, a}(X),
\eeq
where the average gauge and  fermionic fields are denoted, respectively,
 by $A(x)$ and $\psi(x)$, omitting the brackets.
The covariant derivative is $D_\mu \equiv
\del_\mu+igt^a A_\mu^a$ and $F_{\mu\nu}\equiv [D_\mu,D_\nu]/(ig)$.
In the r.h.s., $\eta(X)$ and $j_\mu^a(X)$ are external perturbations; in
particular, they may vanish, in which case Eqs.~(\ref{avpsi})-(\ref{avA})
describe the soft normal modes of the plasma.
The ``induced sources'' show how the collective motion of the hard
particles affects -- through the associated polarization effects -- the
 properties of the mean fields.
They are generally expressed in terms of
connected, off-equilibrium, 2-point functions, and have to be considered as
functionals of the fields themselves. For example,
$\eta^{ind}=g\gamma^\nu t_a\langle A_\nu^a (x)\psi(x)\rangle_c$
involves an abnormal quark-gluon propagator,  which is
 nonvanishing only in the presence of the  fermionic mean field $\psi$.
The induced color current $j_\mu^{ind}$ receives contributions
from various species of hard particles (quarks and gluons).
The fermion piece involves the  quark  propagator in the presence
 of the fields:   $j_{{\rm f}}^\mu=g\,t^a\langle\bar\psi(x)
\gamma^\mu t^a\psi(x)\rangle_c$.
In a covariant gauge, the bosonic piece $j^\mu_{\rm b}$  involve both gluons
and   ghosts\cite{us}.

The 2-point functions which enter the induced sources
are obtained in terms of the mean fields by
solving the appropriate eqs. of motion at the level
of the present approximation (see below Eqs.~(\ref{L})--(\ref{N})).
 It is remarkable that the corresponding
solutions may be expressed in terms of {\it on-shell
 generalized distribution functions for the hard
thermal particles (quarks and transverse gluons).} To be specific, we found
that, in leading order, the induced sources can be written as
\beq\label{jind}
j_\mu^{ind\,a}(X)\,=\,g\int\frac{d^3k}{(2\pi)^3}\,\frac{k_\mu}{k}
\left\{\,N_{\rm f}\left[ \delta
 n_+^a(\vec k,X)-\delta n_-^a(\vec k,X)\right]\,+\,
2N\,\delta N^a(\vec k,X)\right\},\eeq
 \beq\label{eind}
\eta^{ind}(x)\,=\,g\int\frac{d^3k}{(2\pi)^3}\,\frac{1}{k}
\,{\slashchar \Lambda} (\vec k,X),\eeq
where $N_{\rm f}$ is the number of quark flavors.
Here, $\delta n_\pm^a(\vec k,X)$ and $\delta N^a(\vec k,X)$ denote
 oscillations induced by the soft mean fields in the color densities  carried
by the hard quarks, antiquarks and, respectively,  gluons
with momentum $\vec k$. Less familiar,
${\slashchar \Lambda}(\vec k,X)$ represents a
generalized one-body density matrix mixing  fermionic
 and bosonic degrees of freedom.
It carries  fermionic quantum number and
describes fluctuations where, under the action of a soft fermionic mean
 field,  hard quarks are converted into hard gluons and  vice-versa.

All the fluctuations above are determined by the following {\it kinetic
equations}\cite{us}:\newpage
\beq\label{L}
(v\cdot D_X)\slashchar{\Lambda}(\vec k, X)=
-i\,C_f\,(N_k+n_k)\,\slashchar{v}\,\psi(X),\eeq
\beq\label{n}
\left[ v\cdot D_X,\,\delta n_\pm({\vec k},X)\right]=\mp\, g\,\vec
v\cdot\vec E(X)\,\frac{dn_k}{dk},\eeq
\beq\label{N}
\left[ v\cdot D_X,\,\delta N({\vec k},X)\right]=-\, g\,\vec
v\cdot\vec E(X)\frac{dN_k}{dk}.\eeq
Here, $v^\mu\equiv (1,\,\vec v)$, where $\vec v\equiv
\vec k/\epsilon_k$ is the velocity of the hard particle,
$C_f$ is the quark Casimir, $E^i\equiv E^i_at^a$ is the chromoelectric
field, while $\delta N\equiv \delta N^at^a$, etc. In the r.h.s., $N_k$ and
$n_k$ denote, respectively, equilibrium boson and fermion occupation factors.
These eqs. determine  the collective dynamics of
the hard particles in the presence of soft background fields. They exhibit
 many features of classical dynamics, and, in fact, they generalize the Vlasov
eq. to nonabelian plasmas (the covariant line derivative in the l.h.s.
is  the familiar drift term of elementary  kinetic equations). Also note that
{\bf i)} Eqs. (\ref{L})--(\ref{N}) are independent of the gauge fixing
 parameter
which enters calculations in general covariant gauges\cite{us};
{\bf ii)} they transform covariantly under a local gauge
transformation of  the mean fields $A_\mu$, $\psi$ and $\bar\psi$;
{\bf iii)} they are nonlinear in $A_\mu$, because of the
covariant derivatives; {\bf iv)} there is a
remarkable symmetry between the response of the hard quarks
and that of the hard gluons to soft fields.

\section{From kinetic equations to HTL's}

Eqs.~(\ref{L})--(\ref{N}) may be exactly solved for
specified boundary conditions. Then, the induced sources result
 after a supplimentary integration over $\vec k$
 (recall Eqs.~(\ref{jind})--(\ref{eind})). Consider, e.g.,  retarded
 conditions, which is appropiate for a plasma being initially in equilibrium.
The corresponding solutions, which express  the response of
 the plasma to ``soft'' perturbations  vanishing as $X_0\to -\infty$,
may be written as
\beq\label{sole}
\eta^{ind}(X)&=& -i\omega_0^2\int\frac{d\Omega}{4\pi}\,\slashchar{v}
\int_0^\infty dt\,U(X,X-vt)\psi(X-vt),\eeq
\beq\label{solj}
j^{ind}_\mu(X)\,=\,3\,\omega^2_p\int\frac{d\Omega}{4\pi}
\,v_\mu\int_0^\infty dt\, U(X,X-vt)\, \vec v\cdot\vec E(X-vt)\,U(X-vt,X),\eeq
in terms of the parallel transporter  $U(x,y)$  along a {\it straight line}
$\gamma$
  joining $x$ and $y$, $ U(x,y)=P\exp\{ -ig\int_\gamma dz^\mu A_\mu(z)\}.$
The {\it plasma frequencies} are $\omega_0^2\equiv C_fg^2 T^2/8$ and
$\omega^2_p\equiv (2N+N_{\rm f})g^2 T^2/18$; the angular integral runs
over all the directions of the unit vector $\vec v$.

The induced sources (\ref{sole})--(\ref{solj}) show how the mean fields
are  renormalized by their interaction with the hard particles.
Accordingly, they
 act as {\it generating functionals for all (retarded) amplitudes between
soft quasiparticles}.
E.g., the soft quark self-energy  in the presence
of a background gauge field $A_\mu$ is
$\Sigma(X,Y)=\delta\eta^{ind}(X)/\delta\psi(Y)$,
which for $ A_\mu=0$ gives precisely the corresponding HTL:
\beq\label{S}
\Sigma(P) =\omega_0^2\int\frac{d\Omega}{4\pi}\, \frac{\slashchar{v}}
{v\cdot P+i\eta}.\eeq
By further differentiations with respect to $A_\mu$ in (\ref{sole}),
we generate all the HTL's between a quark pair and
any number of soft gluons\cite{us}.
Note that $\eta^{ind}$ is  linear in $\psi$, so there is no
 amplitude with more than one pair of soft  external fermions.
Similarly, all multi-gluon HTL's arise when differentiating
$j_\mu^{ind}(X)$. For instance,
$\Pi_{\mu\nu}(X,Y)=\delta j_\mu^{ind}(X)/\delta A^\nu(Y)$
is the well-known  polarization tensor for soft gluons:
\beq\label{P}
\label{htlpi}\Pi^{ab}_{\mu\nu}(P)=
3\,\omega_p^2\,\delta^{ab}
\left \{-\delta^0_\mu\delta^0_\nu \,+\,P^0 \int\frac{d\Omega}{4\pi}
\frac{v_\mu\, v_\nu} {v\cdot P+i\eta}\right\}.
\eeq
Remark that the denominator $1/(v\cdot P)$ typical for the HTL's reflects
simply the drift operator in the l.h.s. of Eqs.~(\ref{L})--(\ref{N}).

Gauge symmetry implies  a
covariant conservation law for the induced current
\beq \left[D^\mu,\,j_\mu^{ind}(X)\right]=0,\eeq
which leads, by differentiation with respect to the mean fields,
to the simple Ward identities relating the HTL's, alluded to in the
introductory section. For example,
\beq
\Pi_{\mu\nu}\,P^\nu=P^\mu\,\Pi_{\mu\nu}=0,\qquad\qquad\nonumber\\
P_1^\mu\,\Gamma_{\mu\nu\lambda}(P_1,P_2,P_3)=\Pi_{\nu\lambda}
(P_3)-\,\Pi_{\nu\lambda}(P_2),...\eeq
where $\Gamma_{\mu\nu\lambda}$ is the HTL for the 3-gluon vertex.

\section{Effective action for soft  fields}

By eliminating the induced sources from Eqs.~(\ref{avpsi})--(\ref{avA}),
using their explicit expressions (\ref{sole})--(\ref{solj}),
 one obtains nonlinear and nonlocal eqs. of
motion which generalize the Maxwell equations in a polarizable medium.
Note that the quark and gluon modes mix,
as a consequence of gauge covariance.

In general, these eqs.  also include dissipative effects, related to the
Landau damping of the mean fields. This explains the nonvanishing imaginary
 parts of the self-energies (\ref{S})-(\ref{P}),  occuring for
space-like  momenta $P_\mu$  (i.e., when $P_0=\vec v \cdot \vec P$).
However, at the level of the present approximation, there is a
well defined {\it transparency range} where dissipation is absent: this
happens for time-like momenta $P_\mu$, where the Landau damping
is kinematically forbidden. In this range, the
 mean field eqs. of motion are generated by the
 minimal  action principle applied to an effective
action  $S_{eff}=S_0+S_{ind}$.
Here, $S_0$ is the classical QCD action, while $S_{ind}$ contains the effects
of the interactions between
the soft fields and the hard particles:
 $S_{ind}=S_{\rm f}+S_{\rm b}$, with\cite{us}
\beq
\label{Sf}
S_{\rm f}&=& \omega_0^2\int\frac{d\Omega}{4\pi}\int
 d^4 X\int d^4 Y\,
\bar\psi(X)\langle X|\frac
{\slashchar{v}} {i(v\cdot D)}|Y\rangle \psi(Y),
\eeq \beq
\label{Sg}
\qquad S_{\rm b}&=&\frac{3}{2}\,\omega^2_p\int \frac{d\Omega}{4\pi}\int d^4 X
\int d^4 Y \,{\rm tr} \left [ F_{\mu\lambda}(X)
\langle X|\frac {v^\mu v_\nu}{(v\cdot \tilde D)^2}|Y\rangle
 F^{\nu\lambda}(Y)\right ],
\eeq
where $\tilde D\,F_{\mu\nu}\equiv [D,\,F_{\mu\nu}]$,
 and the trace acts on color indices. Note that for undamped fields,
the operator $v\cdot D$ never vanishes, and thus the inverse operators
entering (\ref{Sf})--(\ref{Sg}) are  nonambiguous\cite{us}. The
 action above coincides with the generating functional for HTL's derived
in Refs.\cite{Wong,Braaten92} on the basis of gauge invariance.
One sees from our analysis that $S_{eff}$ may be given the physical
interpretation of the classical action which describes long wavelength
excitations of the hot quark-gluon
plasma, at leading order in the coupling $g$, but only as long as
 Landau damping is inoperative.
 In cases where the boundary conditions matter, (e.g., for computing
 the field energy loss in the plasma),
 we have to recourse to the kinetic equations (\ref{L})--(\ref{N}) and solve
for the induced sources appropriate to the given physical conditions.

\vspace{1cm}
{\bf Acknowledgements}

It is a pleasure to thank the organizers of the Banff workshop, especially
Randy Kobes and Gabor Kunstatter, for their kind hospitality.

\end{document}